\documentclass[11pt]{article}
\usepackage{graphicx}
\usepackage{amsmath}
\usepackage{enumitem}
\usepackage[caption=false]{subfig}
\usepackage[margin=1.25in]{geometry}
\usepackage[usenames,dvipsnames]{color}
\usepackage{url}
\usepackage[colorlinks = true,
            linkcolor = blue,
            urlcolor  = blue,
            citecolor = blue,
            anchorcolor = blue]{hyperref}

\newcommand\pubnumber{}
\newcommand\pubdate{\today}

\def\Title#1{\begin{center} {\LARGE #1 } \end{center}}
\def\Author#1{\begin{center}{ \sc #1} \end{center}}
\def\Address#1{\begin{center}{ \it #1} \end{center}}

\newcommand\pubblock{\rightline{\begin{tabular}{l} \pubnumber\\
         \pubdate \end{tabular}}}
\newenvironment{Abstract}{\begin{quotation} \begin{center}
                       ABSTRACT
     \end{center}\bigskip  }{\end{quotation}}

\def\beq{\begin{equation}}
\def\eeq#1{\label{#1}\end{equation}}
\def\eeqn{\end{equation}}

\newenvironment{Eqnarray}%
   {\arraycolsep 0.14em\begin{eqnarray}}{\end{eqnarray}}
\def\beqa{\begin{Eqnarray}}
\def\eeqa#1{\label{#1}\end{Eqnarray}}
\def\eeqan{\end{Eqnarray}}

\let\bar=\overbar

\def\lsim{\mathrel{\raise.3ex\hbox{$<$\kern-.75em\lower1ex\hbox{$\sim$}}}}
\def\gsim{\mathrel{\raise.3ex\hbox{$>$\kern-.75em\lower1ex\hbox{$\sim$}}}}

\def\del{\partial}
\def\Dslash{\not{\hbox{\kern-4pt $D$}}}
\def\dslash{\not{\hbox{\kern-2pt $\del$}}}
\def\pslash{\not{\hbox{\kern-2pt $p$}}}
\def\ETmiss{\not{\hbox{\kern-4pt $E$}}_T}

\def\Dlr{\mathrel{\raise1.5ex\hbox{$\leftrightarrow$\kern-1em\lower1.5ex\hbox{$D$}}}}

\def\MSB{{\bar{M \kern -2pt S}}}
\def\msb{{\bar{\scriptsize M \kern -1pt S}}}

\def\drb{{\bar{\scriptsize D \kern -1pt R}}}

\newcommand\snowmass{\begin{center}\rule[-0.2in]{\hsize}{0.01in}\\\rule{\hsize}{0.01in}\\
\vskip 0.1in Submitted to the  Proceedings of the US Community Study\\ 
on the Future of Particle Physics (Snowmass 2021)\\ 
\rule{\hsize}{0.01in}\\\rule[+0.2in]{\hsize}{0.01in} \end{center}}

\newcommand{\mytab}{
\begin{tabular}{|c c|}
\hline\hline
Particle & Channel\\
\hline
&$\rightarrow e N_{2,3}$\\
$D_s$ & $\rightarrow \mu N_{2,3}$\\
&$\rightarrow \tau \nu_{\tau}$\\
\hline
&$\rightarrow \nu_{\tau} \mu N_{2,3}$\\
$\tau$ & $\rightarrow \nu_{\mu} \mu N_{2,3}$\\
&$\rightarrow \rho N_{2,3}$\\
\hline\hline
{$N_{2,3}$}&$\rightarrow \pi^+ \mu^-$\\
&$\rightarrow \pi^- \mu^+$\\
\hline
\end{tabular}
}

\begin{document}

\pubblock

\Title{Sensitivity to Heavy Neutral Leptons with the SAND detector at the DUNE ND complex}

\bigskip 

\Author{Zahra Ghorbani Moghaddam\\For DUNE Collaboration}

\medskip

\Address{ }

\medskip

 \begin{Abstract}
\noindent Heavy Neutral Leptons (HNLs) have been an interesting topic for experimental particle physics in the past few years. 
A study has been performed within the framework of the multi-instrument DUNE near detector complex, specifically regarding the on-Axis System, to assess the sensitivity to HNL within six years of exposure.
By utilizing two MC generators, and charmed heavy meson decay channels, the sensitivity to HNL masses between 0.3 and 1.8 GeV/$c^2$ has been explored. A Mad-Graph/Mad-Dump model has been implemented based on the $\nu MSM$ Lagrangian, and used to obtain accurate kinematics for the decay of mesons and HNL.
The simulated final-state particles have been propagated through the detector; a track reconstruction algorithm, based on the Kalman Filter technique, along with a simple two-body decay selection, is implemented to estimate efficiency and background rejection.
The HNL sensitivity has been estimated both from purely phenomenological as well as experimental point of view, reaching O($10^{-9}$) for higher HNL masses, with about a factor 3 deterioration between the phenomenological and the experimental case. 
In this paper, the results for direct and indirect decay channels of charmed meson $D_s$ to HNL has been investigated and the potential for further improvements has been discussed.

\end{Abstract}

\snowmass

\def\thefootnote{\fnsymbol{footnote}}
\setcounter{footnote}{0}
\section{Introduction}
The Standard Model (SM) predictions have been tested and confirmed by numerous experiments, most recently by the Large Hadron Collider (LHC).
Besides the numerous successes of the SM model, no significant deviations in direct or indirect searches for new physics have been observed; yet, exploring and defining ranges and limits for new physics is still on-going.\\
It is clear now that the SM is not a complete framework, as it fails to explain a number of observed phenomena in particle physics, astrophysics and cosmology. These major unsolved challenges are commonly known as, beyond Standard Model (BSM) phenomena;
from the cosmological observations, that lead to the possible existence of dark matter, and the observed matter anti-matter asymmetry of the universe, to recent experimental results on neutrino oscillation\cite{Super-Kamiokande:1998kpq}; all are confirming that active neutrinos have mass, contrary to the predictions of the SM.\\ 
One of the scenarios incorporating such correction into the SM is $\nu MSM$ (neutrino Minimal Standard Model). It introduces a natural renormalizable extension to the SM\cite{Shaposhnikov:2008pf,Gorbunov:2007ak}, and similarly to the See-Saw type I\cite{Minkowski:1977sc,Yanagida:1979gs,Yanagida:1980xy}, it includes three light singlet RH fermions or Heavy Neutral Leptons (HNL), as the extended fields.
The leptonic sector of the theory has the same structure as the quark sector, i.e. for each left-handed fermion there is a right-handed counterpart. This model is not only consistent with the data on neutrino oscillations but also provides a candidate for dark matter particle, the lightest singlet fermion (HNL) amongst the three. The $\nu MSM$ scenario includes the baryon asymmetry of the Universe (BAU) as well \cite{Gorbunov:2007ak}, which depicts an elegant picture in explaining three BSM phenomena in one theoretical framework.
The three RH Majorana neutrinos added to the SM Lagrangian are the low-energy representation of LN-violating Weinberg operator, $\mathcal{L}_5\propto\frac{C^{ll'}}{\Lambda}[\phi.\bar{L}^c][L.\phi]$ \cite{Keung:1983uu}. Since the energy scale of HNLs in this model is much lower than the electroweak scale, it offers possibility for direct searches. 
From the experimental point of view, heavy neutral leptons (HNLs) with mass range below the beauty meson (B) have higher chances for detection. There are various experiments, from LHC to beam-dump facilities, in pursuit of HNLs, like SHiP\cite{SHiP:2015gkj}, FASER\cite{FASER:2018eoc}, MATHUSLA\cite{MATHUSLA:2018bqv} and NA62\cite{NA62:2017rwk,NA62:2020mcv}. Some neutrino experiments in the beam-dump mode are also active in search for the HNLs, like T2K\cite{T2K:2019jwa} and DUNE\cite{DUNE:2018tke, Ballett:2019bgd}.
DUNE Near Detector (ND) has promising capabilities to improve the existing HNL sensitivity constraints in the mass-coupling phase space\cite{Ballett:2019bgd}.

DUNE, with its 1300 km baseline, is the largest long-baseline neutrino experiment. Each Far Detector (FD) modules consists in 17 kT Liquid Argon Time Projection Chambers (LArTPCs)\footnote{The total active volume to detect neutrino interactions is 40 kiloton}, located 1.5 km underground, at Sanford Underground Research Facility. The Near Detector (ND) is a hybrid design based on three different detectors, two of which with the ability to go off-axis, and it is placed 62 m underground, at a distance of 574 m from the target\cite{DUNE:2018tke}.
The beam facility produces the neutrino beam by extracting protons from the Fermilab Main Injector. The proton energy is in the range 60-120 GeV, at a nominal power of 1.2 MW, possibly upgraded to 2.4 MW. Approximately $7.5\times 10^{13}$ protons is extracted every 1.2 seconds at 120 GeV, for an estimated total number of $1.11 \times 10^{21}$ pot/year (NPOT).
The DUNE ND complex design consists in, LArTPC, muon-spectrometer and a System for on-Axis Neutrino Detection (SAND)\cite{DUNE:2018tke, DUNE:2021tad}. 

SAND is a permanently on-axis detector with magnetized tracker. The magnet\cite{Smith:1997qhd} is designed in conjunction with its iron yoke to produce 0.6 T over a 4.3 m length, 4.8 m diameter volume, enveloping the multi-target tracker, Straw Tube Tracker (STT), and LAr meniscus at the upstream end\cite{DUNE:2021tad}. 

\section{Model Specifications}
The $\nu MSM$ Lagrangian includes three heavy right-handed fields, $N_{1,2,3}$, where the lightest, $N_1$, represents the dark matter candidate, and the others two are responsible for baryogenesis and generation of lepton asymmetry through resonant production of $N_1$\footnote{This theory can accommodate inflation, if one considers non-minimal coupling between the Higgs boson and gravity\protect\cite{Shaposhnikov:2006xi}}\cite{Gorbunov:2007ak}.
\begin{equation}
\mathcal{L}_{\nu MSM} = \mathcal{L}_{MSM} + \bar{N}_I i \partial_{\mu}\gamma^{\mu} N_I - F_{\alpha I}\bar{L}_{\alpha} N_I\Phi - \frac{M_{IJ}}{2}\bar{N}^c_I N_J + h.c.
\label{eq:Lag}
\end{equation}
I and J are corresponding to the flavor indices of HNLs and run from 1 to 3. The mass range of sterile neutrinos in this model is of the same order of the masses of other leptons in the SM, O(KeV-GeV). In order to have quantitative predictions, three sets of Yukawa coupling benchmarks can be defined for three extreme hierarchies\cite{Gorbunov:2007ak,Shaposhnikov:2008pf}.
\begin{equation}
\begin{aligned} 
\text{Model I   :  } f^{2}_{e}\hspace{0.4em} : \hspace{0.4em}f^{2}_{\mu} \hspace{0.4em} : \hspace{0.4em}f^{2}_{\tau} \hspace{0.4em} &\approx  \hspace{0.4em} 52    \hspace{0.4em}:  \hspace{0.4em}1  \hspace{0.4em} : \hspace{0.4em}1 \\ 
\text{Model II  :  } f^{2}_{e}\hspace{0.4em} : \hspace{0.4em}f^{2}_{\mu} \hspace{0.4em} : \hspace{0.4em}f^{2}_{\tau} \hspace{0.4em} &\approx  \hspace{0.4em} 1     \hspace{0.4em}:  \hspace{0.4em}16 \hspace{0.4em} : \hspace{0.4em}3.8 \\ 
\text{Model III :  } f^{2}_{e}\hspace{0.4em} : \hspace{0.4em}f^{2}_{\mu} \hspace{0.4em} : \hspace{0.4em}f^{2}_{\tau} \hspace{0.4em} &\approx  \hspace{0.4em} 0.061 \hspace{0.4em}:  \hspace{0.4em}1  \hspace{0.4em} : \hspace{0.4em}4.3 \\ 
\end{aligned}
\end{equation} 
To simulate the model, the Lagrangian and its parameters have to be translated or built by FeyRules\cite{Degrande:2011ua}. The Universal FeynRules Output (UFO) contains sets of Python modules with the particle's specifications, making the output flexible to link with other software packages. Since the dominant HNL generation channel is through hadron decay\footnote{The generated HNL from other sources, like the interaction of the $\nu$ beam in the rock, show noncompetitive with respect to the hadron decay. Based on a toy MC study the large dispersion angle of the generated HNLs makes the acceptance for such sources highly suppressed}, the model made by FeyRules involves only charmed meson decay\footnote{dominant in HNL generation up to 2 GeV} to HNL via effective vertices\protect\cite{Ebert:1994tv,Kaplan:1998we}.

\section{Simulation}

Concerning the meson decay channels, the spectrum of the outgoing HNL in a given experiment is determined mostly by the spectrum of produced hadrons, decaying directly, or indirectly, via an intermediate channel, to HNL. The number of produced hadrons can determine the HNL flux. Since the charmed mesons, $D_s$ in particular, dominates the HNL generation, in the mass region up to 2 GeV\cite{Ballett:2019bgd, DUNE:2018tke}, this work covers the dominant direct and indirect channels of $D_s$. 

The MC simulation works in two parts. The first part generates the $D_s$ flux through Pythia8\cite{Sjostrand:2014zea} from a simulated DUNE proton beam. The DUNE proton beam of 80-120 GeV hits graphite target of 2 m, producing mesons, whith the heavy ones decaying inside the target\cite{DUNE:2018tke}. 
Pythia8 is responsible to simulate $10^7$ p-p interaction with 120 GeV in beam-dump mode\footnote{In order to improve the generation efficiency, $D_s$ has been generated by forcing the $c\bar{c}$ interaction and then the result is normalized to the least biased setup}.\\
The second part of the simulation is carried out by Mad-Dump2.0\cite{Buonocore:2018xjk}, which is a minimally modified Mad-Graph5\_aMC@NLO for fixed-target experiments. One of the advantages of using Mad-Dump over Mad-Graph is its efficient flux generator that keeps into account spatial correlations and the geometry of the experiment.

\begin{figure}[!htbp]
\begin{minipage}{.5\linewidth}
\centering
\subfloat{\mytab}%
\caption{\small{HNL production and decay channels included in the single event sensitivity estimate.}}%
\label{fig:channel}
\end{minipage}%
\begin{minipage}{.8\linewidth}
\includegraphics[width=0.6\textwidth]{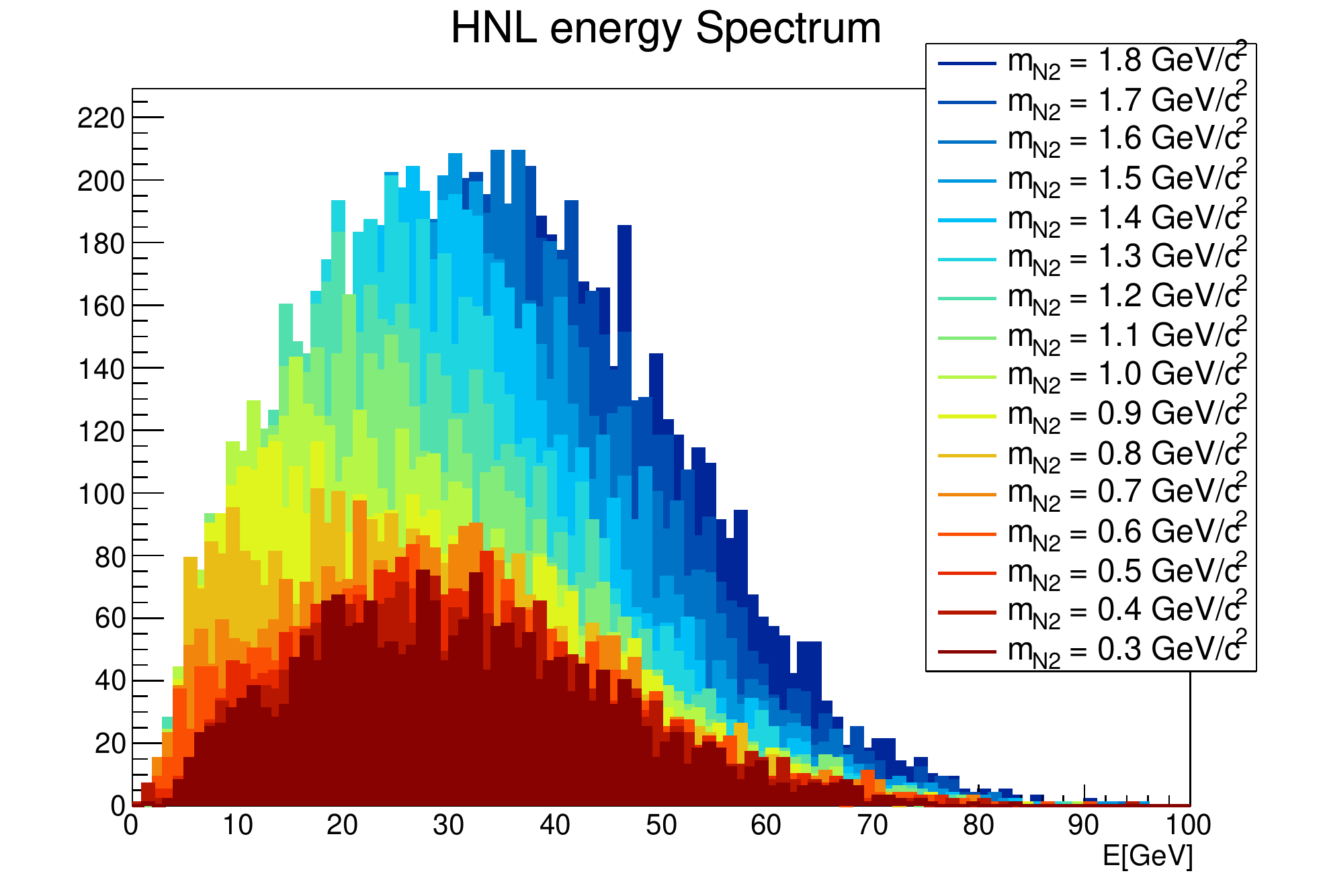}
\includegraphics[width=0.6\textwidth]{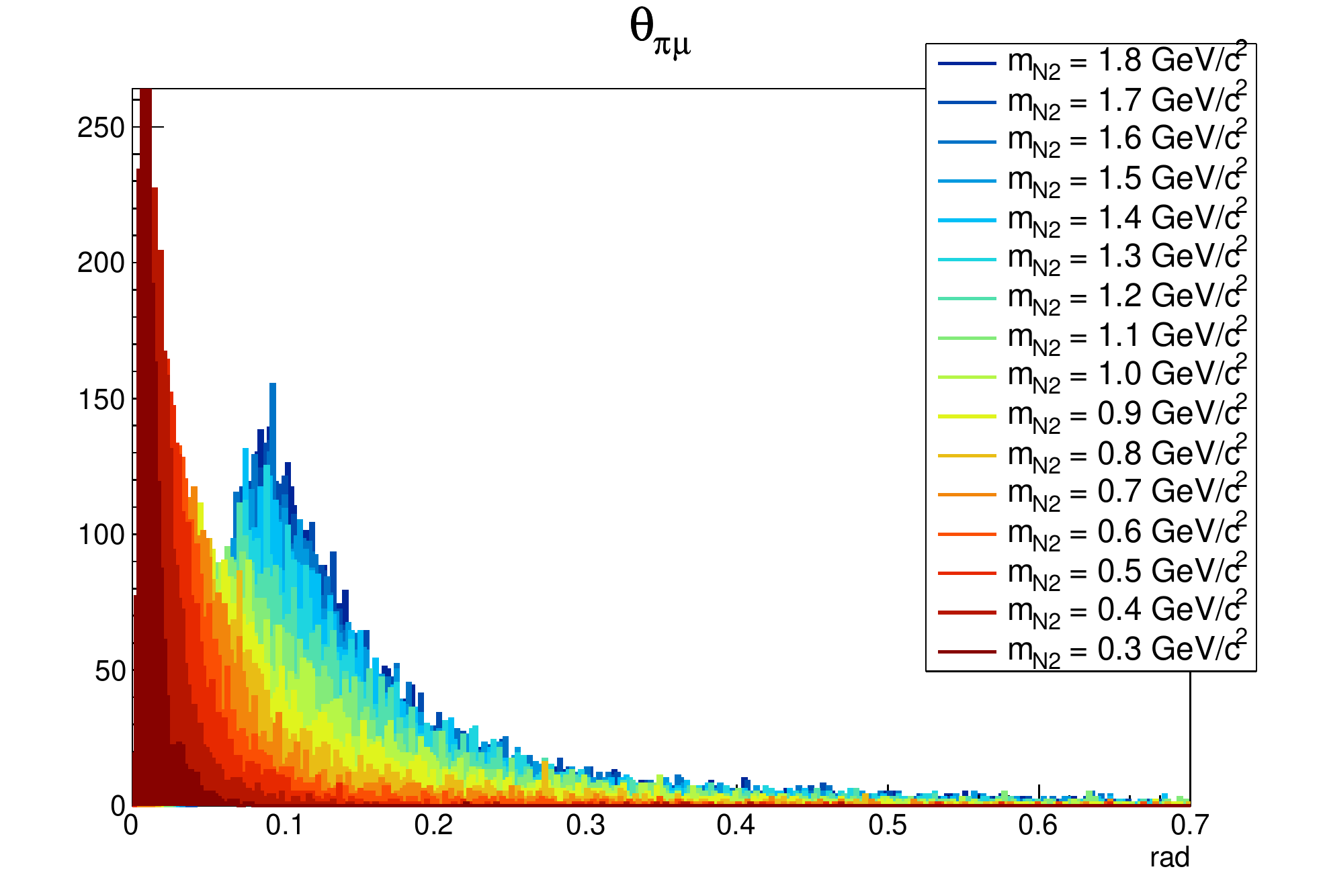}
\caption{\small HNL energy spectrum (top), and\\ opening angle of the final state (bottom) of\\ $D_s\rightarrow \mu N_2$ and $N_2\rightarrow \pi \mu$.}
\end{minipage}%
\end{figure}

The smoking gun signal for HNL in LHC experiments is the di-lepton signal, through which the nature of heavy neutrinos can be determined as well, due to the allowed lepton number violating (LNV) processes for Majorana-like HNL\footnote{Same lepton charge (LNV) for sibling and daughter leptons in the Majorana-like HNL mediated processes}. However, since the sibling lepton in fixed-target experiments is not detectable (hadrons decay outside the detector), the signal relies on the displaced decay of long-lived HNL inside the detector.
Which decay channel of the HNL to observe also matters when it comes to detection. The charged current channels are usually better signals amongst which the channels with $\mu$, because of its clean signature, are usually preferred.  
In order to improve the efficiency of the MC simulation, the displaced vertex signal is forced inside the target detector, SAND, and a weight factor is calculated for each event to compensate for such assumption.
\subsection{Single Event Sensitivity}
The Single Event Sensitivity (SES) is the estimate of minimal coupling ($U_i^2$) for which at least one event can be observed. Estimating the number of observed signal events requires the combination of all cross sections and branching ratios related to the chain of processes, starting with the beam interaction in the target and ending with the final state in the detector (see the table in Fig.\ref{fig:channel}). 
\begin{equation}
N_s \hspace{0.2em}=\hspace{0.2em} \mathcal{C}\displaystyle\sum_{h,f}{N(pp\rightarrow h)BR(h\rightarrow N)(U^2)BR(N\rightarrow f)(U^2)}\displaystyle\sum_{i=ibin}{\mathcal{W}(\gamma\beta[i]\times\gamma\beta[i])}
\label{eq:NS}
\end{equation}
$N_s$ is the number of observed events, here equal to 1; $\mathcal{W}$ stands for the weight factor for each event, and $\mathcal{C}$ is a constant factor including exposure time, NPOT, and normalizations\footnote{all normalizing factors have been taken into account conservativly}. 
\begin{figure}[!htbp]
    \centering
    \subfloat[]{\includegraphics[width=0.55\textwidth]{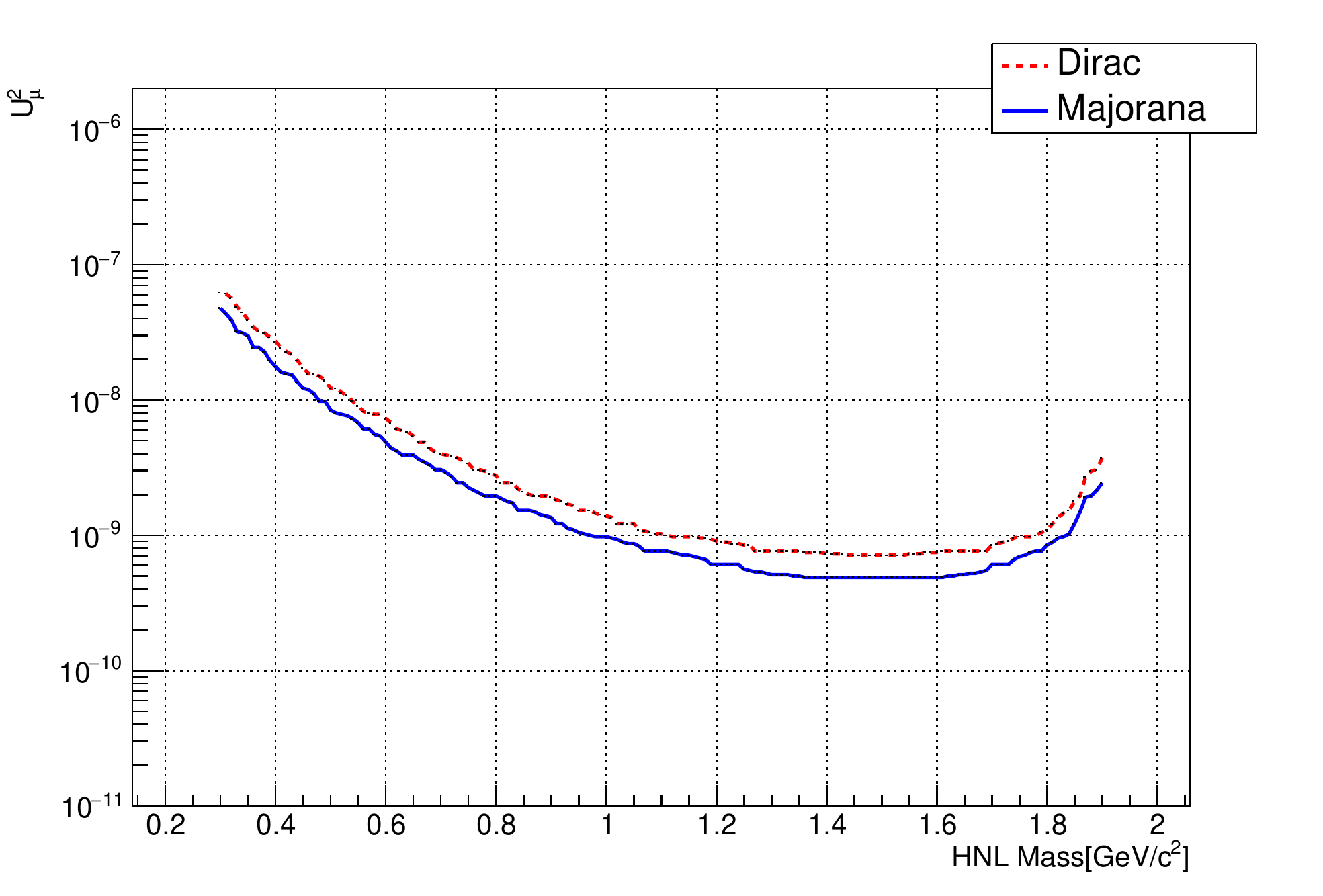}}\\
    \subfloat[]{\includegraphics[width=0.55\textwidth]{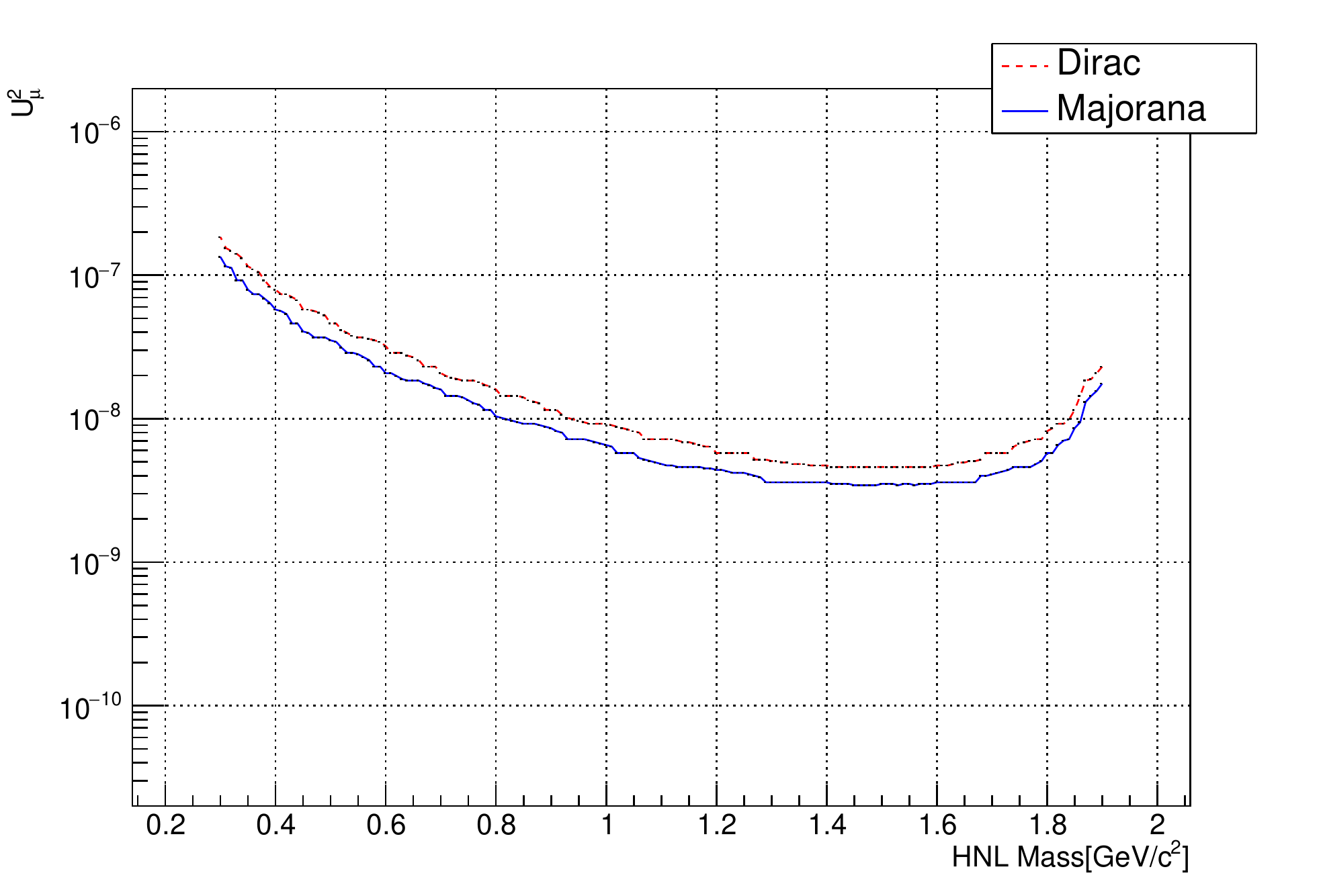}}\\
    \subfloat[]{\includegraphics[width=0.55\textwidth]{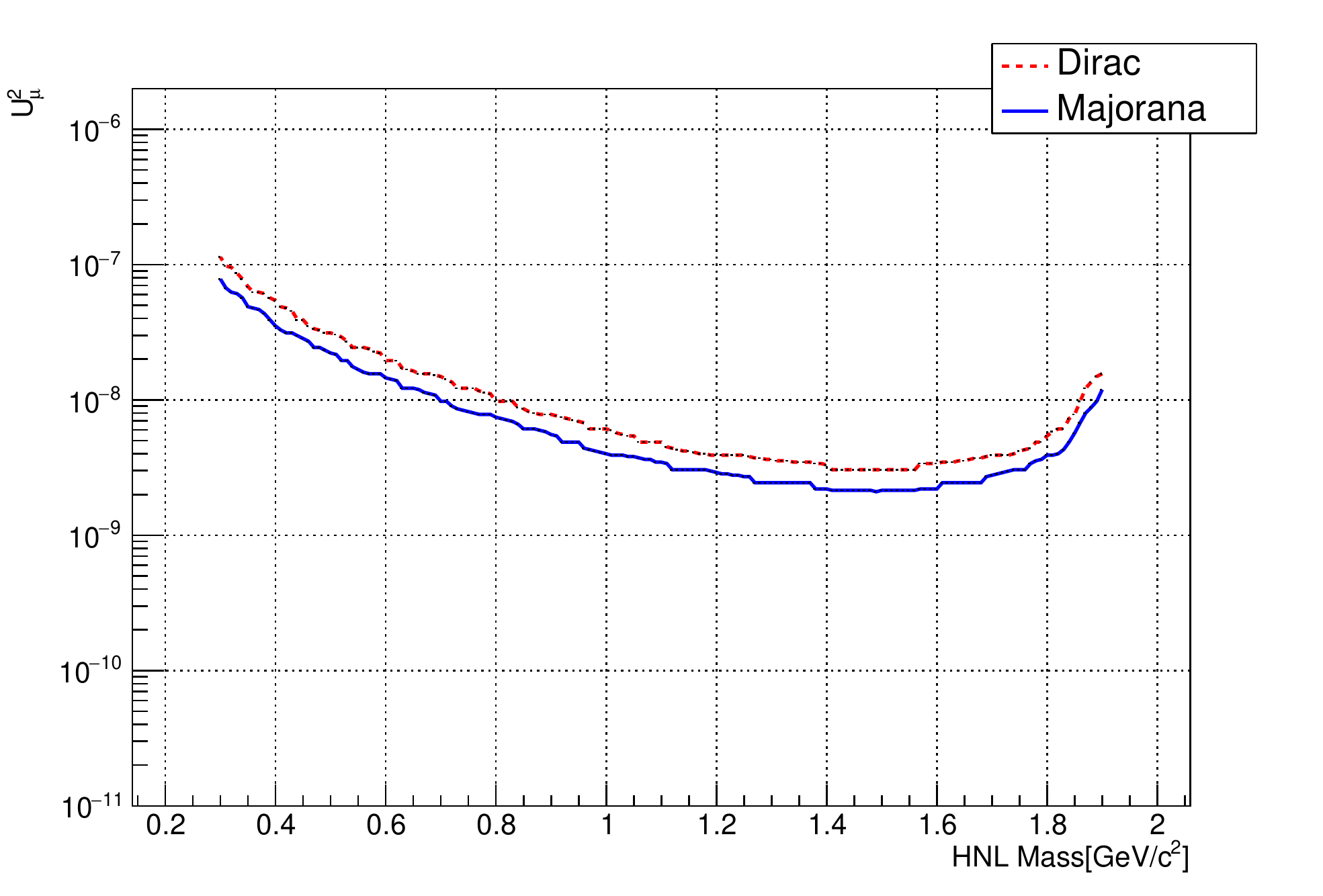}}
    \caption{\small The $U^2_{\mu}$ sensitivity to all direct and indirect channels of $D_s$ for six years of exposure showing for three benchmark models. (a) $U_{e}:U_{\mu}:U_{\tau} \sim 52:1:1$, (b) $U_{e}:U_{\mu}:U_{\tau} \sim 1:16:3.8$, (c) $U_{e}:U_{\mu}:U_{\tau} \sim 0.061:1:4.3$}
    \label{fig:sestot}
\end{figure}

Fig. \ref{fig:sestot} shows the phenomenological sensitivity for three benchmark models within the context of $\nu MSM$. The covered channels are $D_s$ direct and indirect decays to HNL, with $\pi\mu$ final state, assuming 100$\%$ reconstruction efficiency and no background. Such result is compatible with other works, in the $D_s$ mass region\cite{Ballett:2019bgd}.

\section{Reconstruction}

To start with the detector simulation, the generated flux by Mad-Dump needs to get converted into a GEINE-like format\footnote{Neutrino experiments use GENIE\protect\cite{Andreopoulos:2009zz} as neutrino event generator. Any other output should be converted to GENIE format to be compatible with the existing detector simulation}, so as to fit in the existing processing chain: 
\begin{equation*}
\text{\textbf{Model}} \Rightarrow \text{\textbf{MC-Gen}} \Rightarrow \text{\textbf{Det-Res}} \Rightarrow \text{\textbf{Digit}} \Rightarrow \text{\textbf{Reco}}
\end{equation*}
In the detector simulation step, the software is using a GEANT4 interface, EdepSim, to process the events for a given geometry, STT in this case. The Edepsim output is, then, digitized with 200$\mu$m smearing, compatible with the detector's resolution.
In order to proceed with the reconstruction of the HNL flux, a customized Kalman Filter (KF) algorithm has been developed, tailored to the needs of the case at hand.
KF is a recursive algorithm that can be applied to any dynamical system, and takes into account the gaussian fluctuations. It estimates the trajectory of the system's state vector, given the observed measurements. It proceeds progressively from one measurement to the next, performing a least square (LS) fit, and updates the information on the state vector's trajectory with each new measurement. KF is extensively used not only in particle track fitting, but also for positioning and navigating purposes, like tracking a ballistic object on a radar screen\cite{Mankel:2004yv}.
KF algorithm is made of three major operations: Predict, Update/Filter, Smooth; each of which applies several matrix operations on the state vector and on its corresponding covariance matrix. For an efficient track finding, the choice of the track parameters of the state vector is paramount. The parameters should be chosen in a way not only to have the least dependence to one another, but also to simplify the calculations.
Similarly, the other matrices, like transport and projection should be chosen to avoid correlations between the parameters, and also to be as linear as possible\cite{Mankel:2004yv}.

\begin{equation}
X =(x,y, t_{x}, t_{y}, \frac{q}{P_{T}}) \hspace{5em}
\frac{q}{P_{T}}[\frac{e}{GeV}]=\frac{1}{R\times 0.3 \times B}[\frac{1}{cm\hspace{0.5em}T}]
\label{eq:statevector}
\end{equation}
\setcounter{footnote}{0}

The parameters of the state vector $X$ are chosen to achieve a stable and converging KF for the geometry at hand\footnote{For example one could chose total momentum instead of $P_T$; however, $P_T$ shows more stability during KF runs}. 
$B$ is the magnetic field, set to 0.6 T to be compatible with the SAND magnet, $R$ is the radius of curvature, and $P_T$ is the transverse momentum of the particle to the magnetic field (B).
Besides the measurement errors, the parameters of the propagated state vector (track) can be affected by random perturbations from the interaction of the particle with the detector material.
These interactions can result in kinks in the particle's trajectory, by which both direction and the energy of the particle is changed consequently. The KF technique includes an error matrix, Q, which provides an effective method to deal with these perturbations\cite{Mankel:2004yv,Billoir:2021srr}. In this case the Q matrix has minimal impact on the particle trajectory due to the high momentum of the particles and the low mass of the detector\cite{Sergi:2012twa}. 
To increase the quality of the reconstructed tracks, forward and backward directions have been implemented to KF passes\footnote{Inspired by Ranger track fitting\protect\cite{Mankel:2004yv}}. Customized for this work, the KF runs under the following assumptions:
\begin{enumerate}
\setlength{\itemindent}{+.1in}
\item Constant, uniform magnetic field of 0.6 T.
\item Discreteness. Detector layers should be in exact z coordinate (no uncertainty on the z coordinate of the detector planes).
\item Forward and Backward passes. Second pass of the Kalman is independent from the first pass, using the same input.
\item Merging Forward and Backward reconstructed tracks, based on 50$\%$ shared hits.
\item Implementation of an external helical fit to optimize the resolutions, invariant mass and momentum (in the limit $Q\ll 0$). 
\end{enumerate}

\begin{figure}[!htpb]
  \centering
  \includegraphics[width=1.1\linewidth]{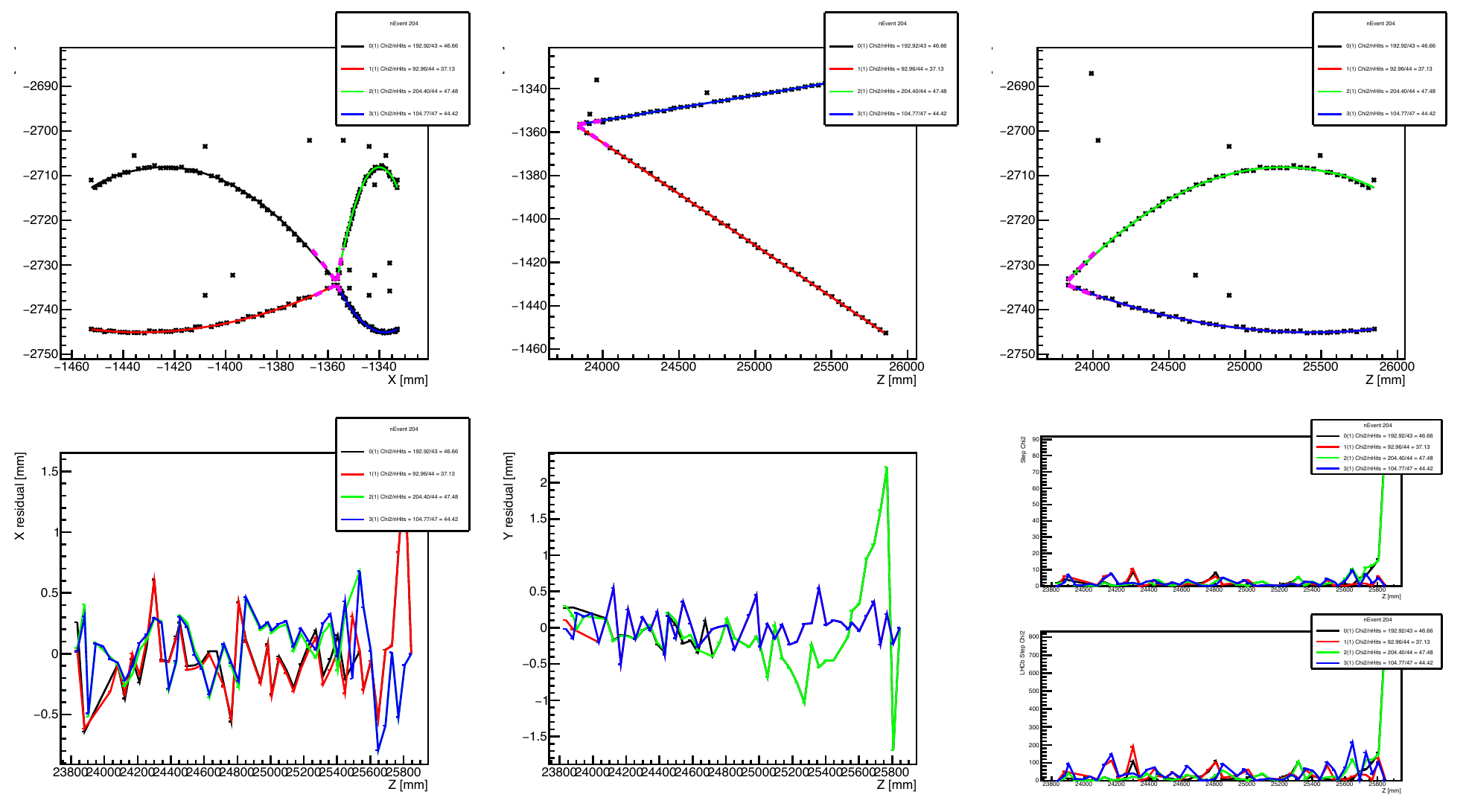}
  \includegraphics[width=1.1\linewidth]{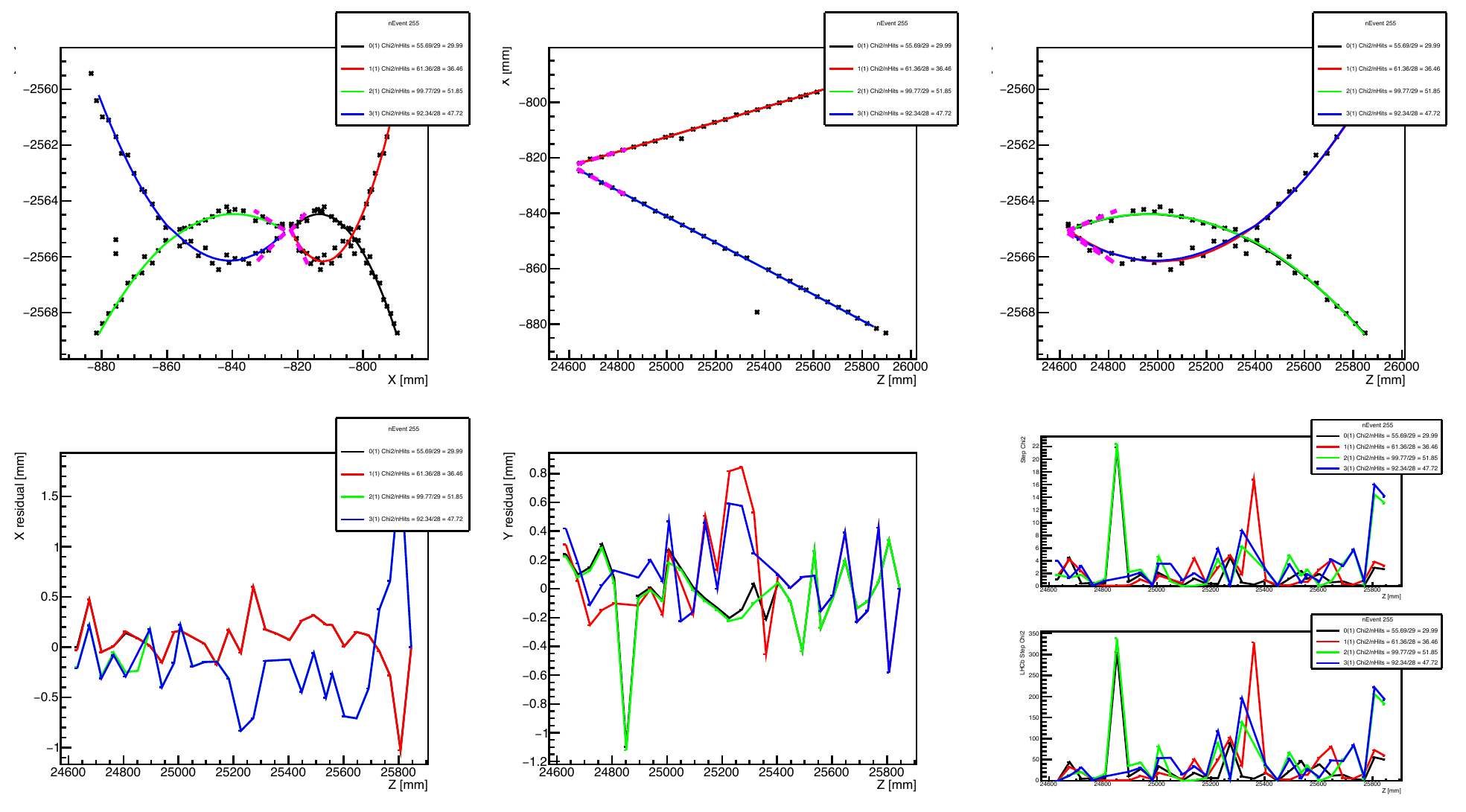}
  \caption{\small{Different views (YX: top left, XZ: top middle, YZ: top right) of two GEANT4 event displays, reconstructed by Kalman Filter. In one event, top plots are different views and the bottom plots are the X and Y residuals, helpful in visual inspection of the KF performance. The bottom right plots are corresponding to different $\chi^2$ schemes. The threshold cut for rejection $\chi^2|^c_{update}\sim 30$.}}
  \label{fig:ExampleEvent}
\end{figure}

Fig. \ref{fig:ExampleEvent} shows different views of two reconstructed events. The black dots are the hits from the digitization and the solid lines are the reconstructed tracks; the two implemented $\chi^2$ schemes are responsible for cleaning up the events from the outlier hits.

Due to the geometry of the STT modules, the way in which the X and Y coordinates of the hits are extracted separately and then recombined results in each event presenting not only two reconstructed tracks, but also two ghost tracks. The coordinate combinatorial $(X,Y)$ results in mirroring the event and creating two ghost tracks in addition to the physical ones. The $\chi^{2}$ is not much of a discriminant when it comes to the distinction of the ghost tracks from the physical ones (e.g. in mirrored events). Nevertheless, the ghost tracks can be, in principle, rejected using the event's kinematics and geometrical features.\\
The efficiency of the KF is approximately 80$\%$ for single tracks and 60$\%$ for a pair of tracks.

\subsection{Selection and Ghost Treatment}

\begin{figure}[!htbp]
    \centering
    \subfloat[]{\includegraphics[width=0.5\textwidth]{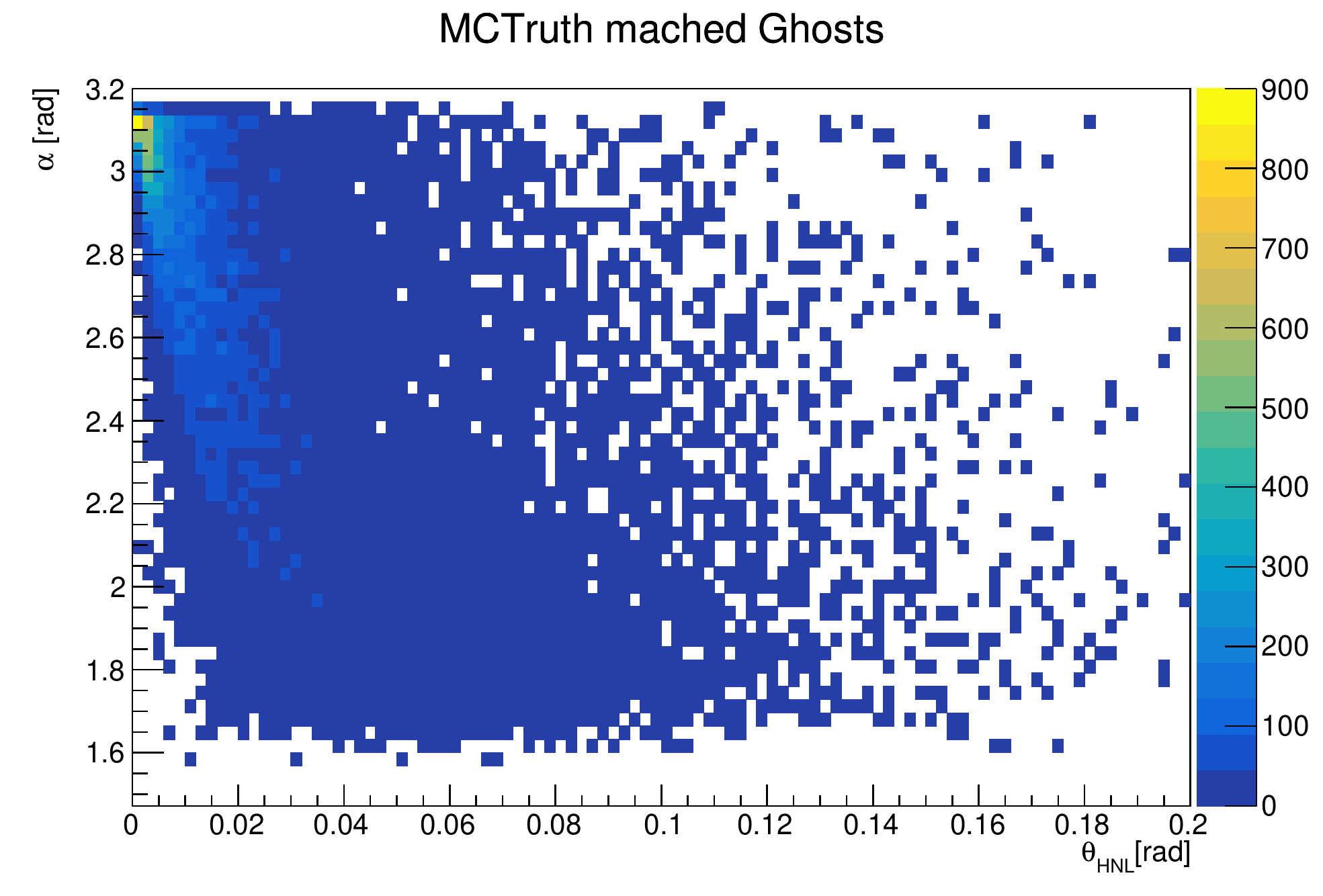}}
    \subfloat[]{\includegraphics[width=0.5\textwidth]{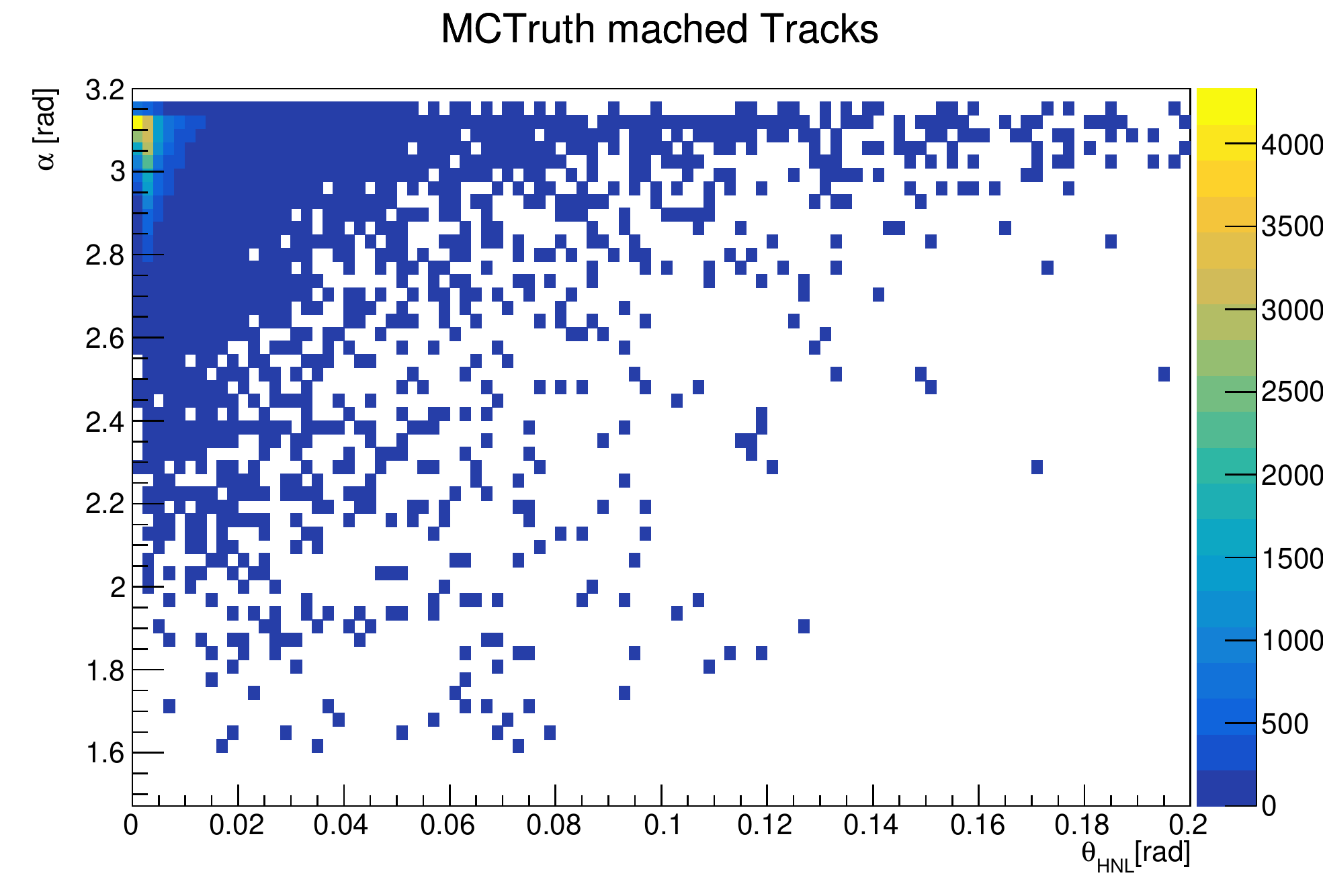}}\\
    \subfloat[]{\includegraphics[width=0.5\textwidth]{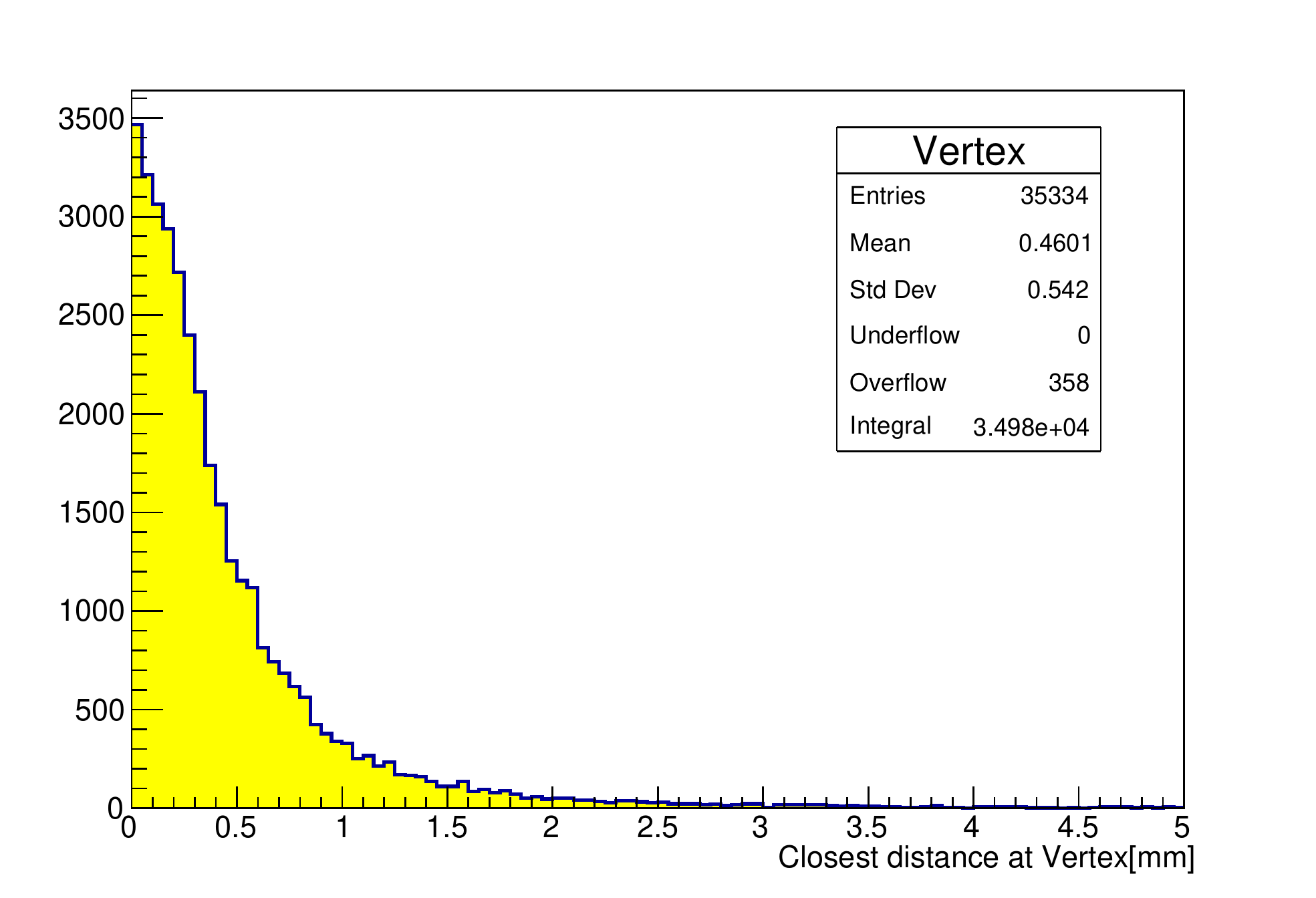}}
    \subfloat[]{\includegraphics[width=0.5\textwidth]{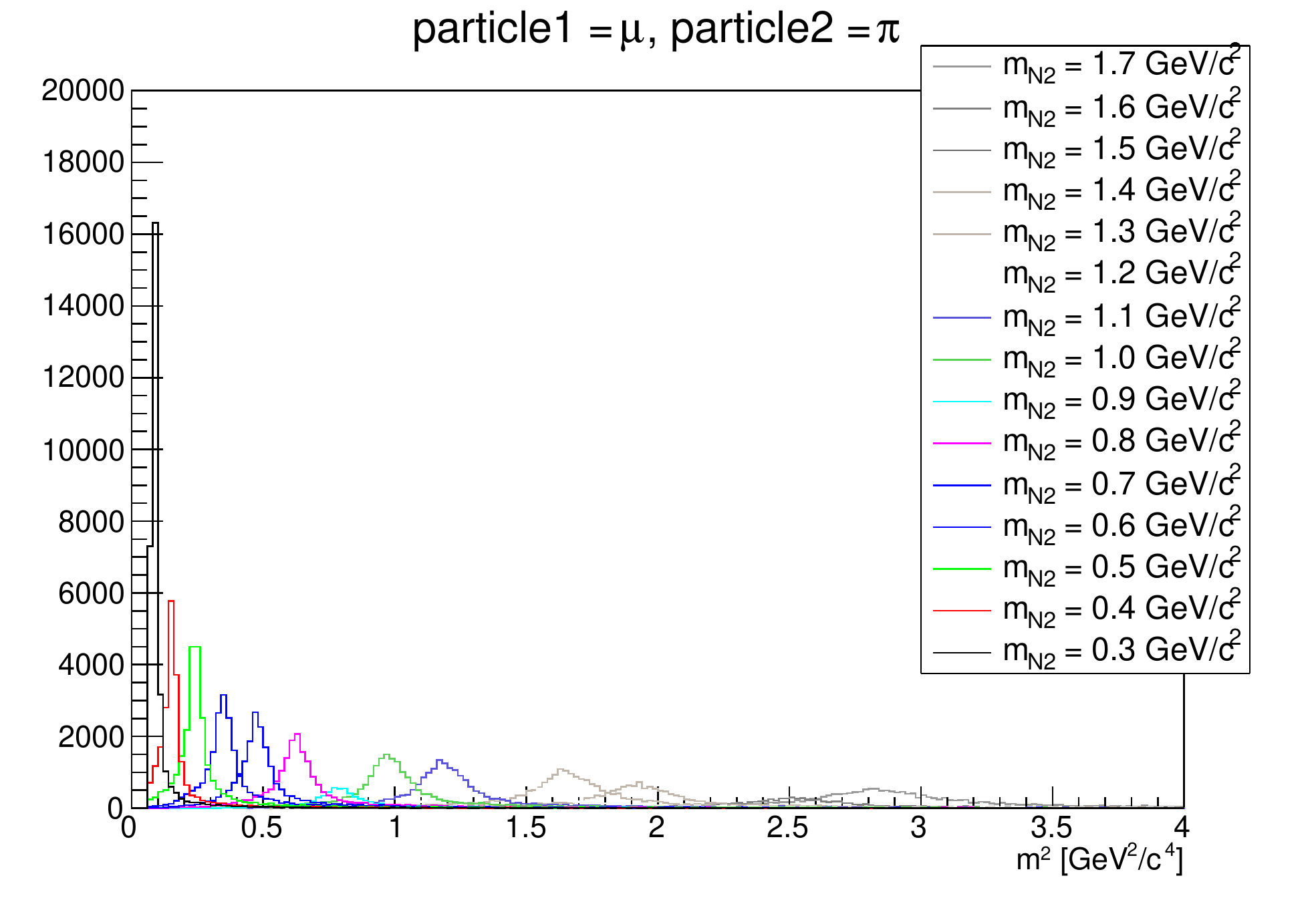}}
    \caption{\small{MC truth matched tracks for direct channel $D_s\rightarrow \mu N_2$. Contributing variables to the selection and potential background rejection schemes: (a) the angle between the ghost tracks, (b) the angle between reconstructed (physical) tracks; (c) shows the minimum distance between the two reconstructed tracks; (d) the invariant mass before the selection.}}
    \label{fig:mcmatched}
\end{figure}

In order to isolate physical tracks from the ghosts to build the signal candidates, a selection scheme, based only on reconstruction information, is needed.
For a two-body decay, signal candidates are composed of two accepted and reconstructed tracks; kinematics and geometry can be exploited to reject unwanted track pairs.
Since the kinematics of a two-body decay imposes the decay products to be back to back in the rest frame, with a significant boost along the $z$ axis, it results in a nearly back to back configuration in the $x-y$ plane. This combination can help in distinguishing the ghosts from the physical tracks in a reconstructed event. 
Fig. \ref{fig:mcmatched} (a) and (b) show the relation between $\alpha$, the opening angle of the tracks, and $\theta$, the angle with respect to the z-axis, for both physical and ghost tracks. It can be seen that by introducing a cut, most of the ghosts can be discarded. 
The selection can also benefit from a cut on the minimum distance between the two tracks, vertex point, see Fig. \ref{fig:mcmatched} (c).\\ 
The selection criteria can be summarized as follows:
\begin{itemize}
\setlength{\itemindent}{+.1in}
\item Tracks in opposite quadrants.
\item Tracks with opposite charge.
\item Angle cut in $\alpha-\theta$ plane, $\alpha > 2.9$ and $\theta<0.02$, see Fig. \ref{fig:mcmatched}.
\item Vertex (the shortest distance between the two reconstructed tracks) $< 1mm$.
\end{itemize}

\begin{figure}[!htbp]
    \centering
    \subfloat[]{\includegraphics[width=0.5\textwidth]{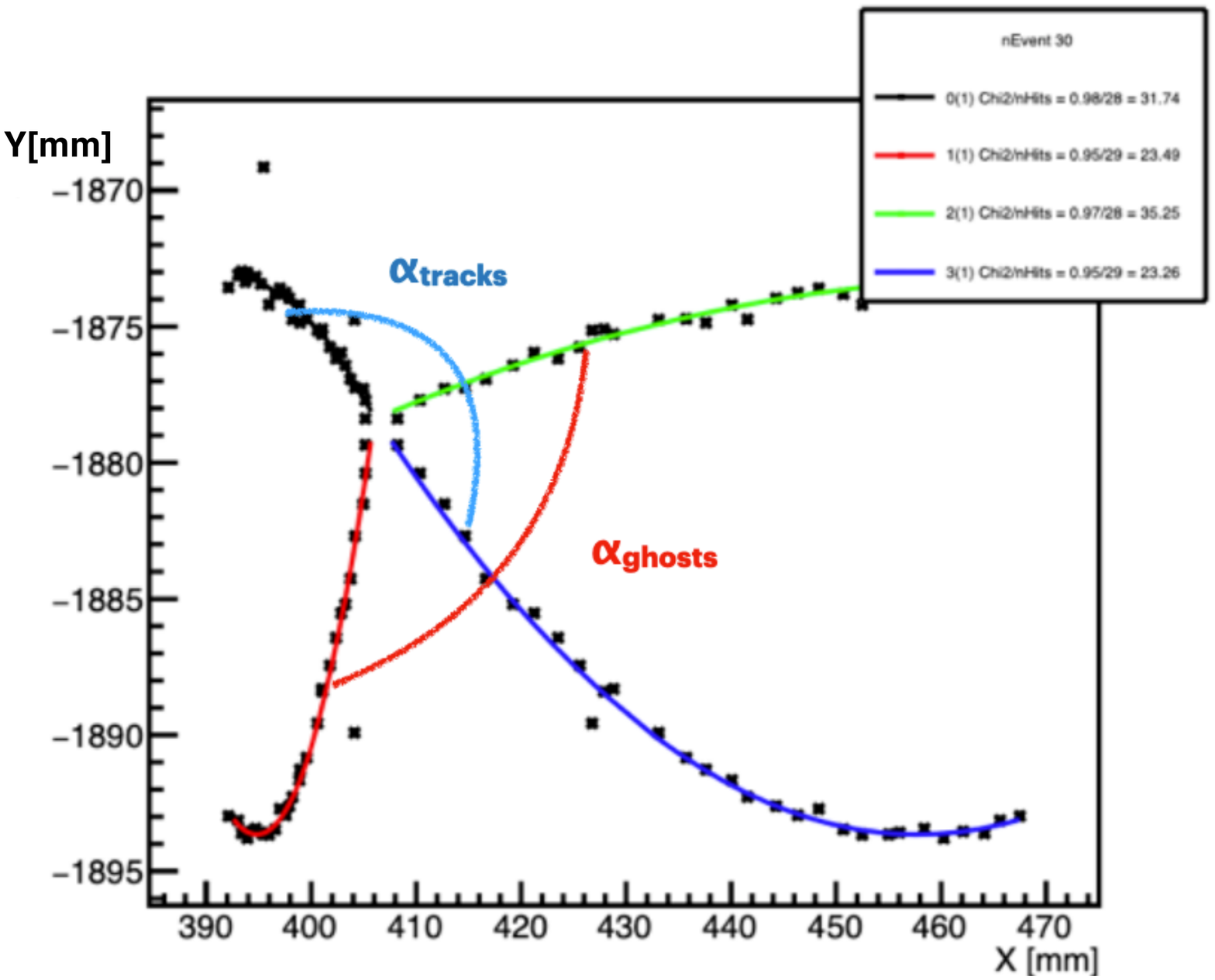}}
    \subfloat[]{\includegraphics[width=0.5\textwidth]{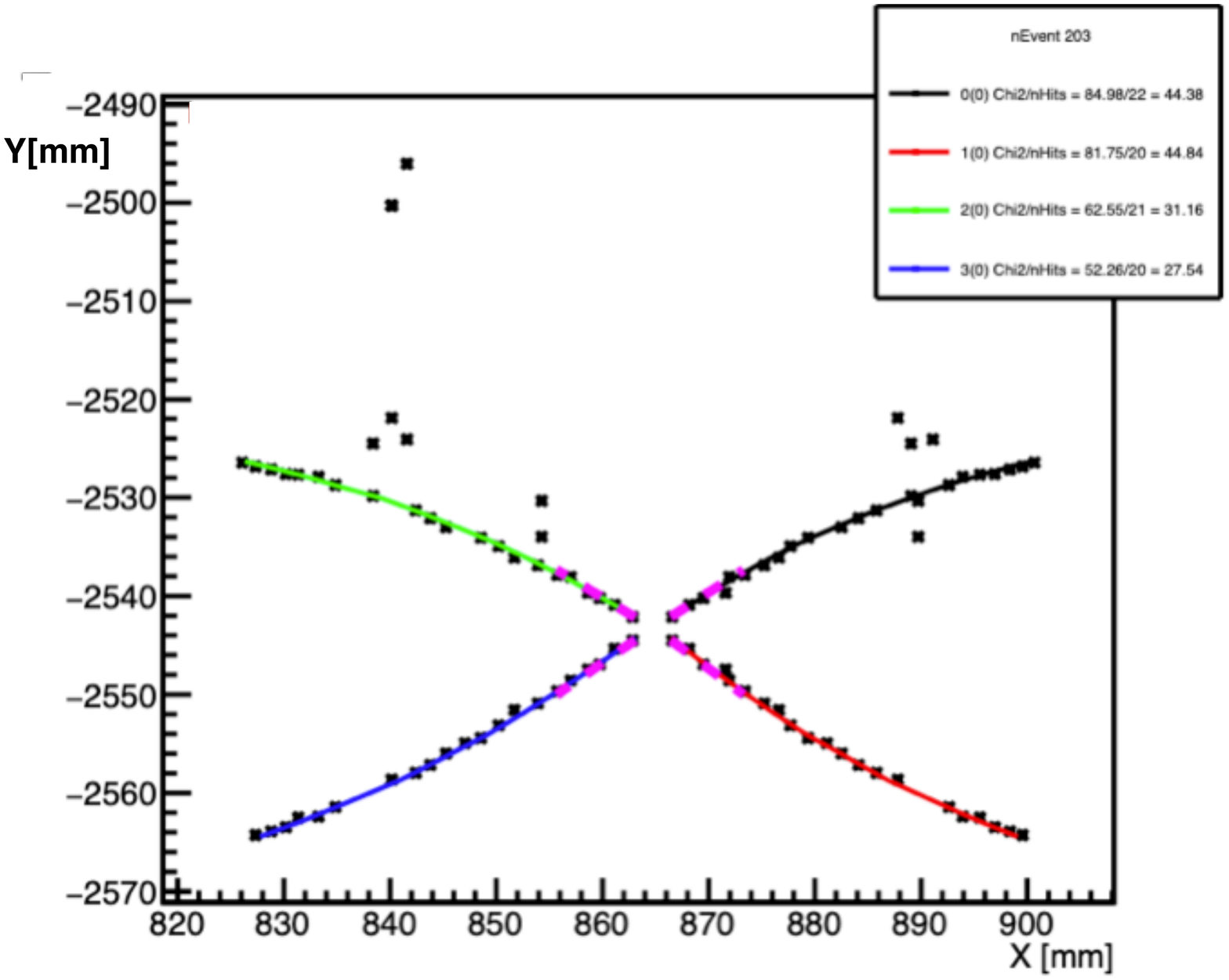}}
    \caption{\small{An example of (a) Asymmetric and (b) Symmetric event, $\alpha_{tracks}$ is the angle between reconstructed tracks, and $\alpha_{ghosts}$ is the angle between the ghosts. In symmetric events, ghosts seem to have no effect on resolutions (momentum and invariant mass); in asymmetric events, however, $\alpha$ allows the distinction between the physical and the ghost tracks}}
    \label{fig:kalmanghost}
\end{figure}

Fig. \ref{fig:kalmanghost} shows events with physical and ghost tracks; (a) demonstrates an asymmetric event, in which the angle between the tracks and ghosts are different, while (b) shows a fully symmetric, mirrored event, where the tracks and ghost are indistinguishable. The ghost contamination coming from symmetric events will be compensated by a correction factor. 

\subsection{Acceptance and Efficiency}
From all the tracks reaching the detector, the long ones within the fiducial volume are accepted. 
If the extrapolation of the tracks to the detector walls includes 6 or more traversed planes (each plane is one STT module), the track is considered long and accepted, otherwise rejected.	
Consequently, an accepted event includes two true particles inside the fiducial volume, amongst the events with 4 reconstructed tracks. Furthermore, to reject badly reconstructed tracks, a $<1$ mm cut on the vertex residual can also be applied. The fraction of accepted events over the total number of generated ones in a cubic fiducial volume defines the efficiency, which is different for each channel. The efficiency estimate shows high sensitivity to the geometry of the event; when the angle between the two candidate tracks is too small, i.e. in the case of HNL mass close to the final state production threshold and therefore low transverse momentum, the reconstruction struggles to separate the two tracks, resulting in efficiency. The efficiency is also showing an inverse relation to the fraction of ghost contamination.
\subsection{Signal Modelling}

\begin{figure}[!htbp]
    \centering
    \subfloat[]{\includegraphics[width=0.55\textwidth]{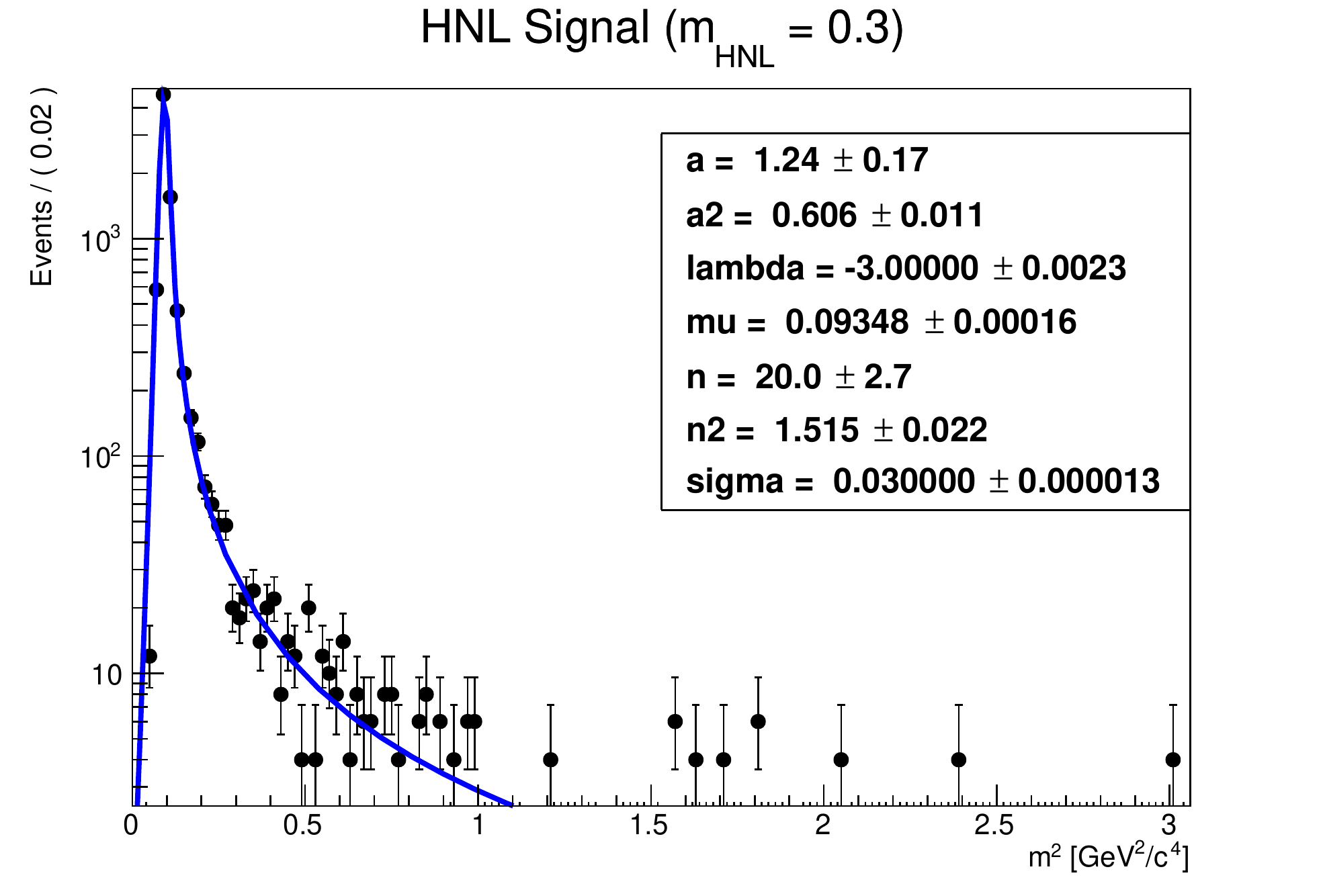}}
    \subfloat[]{\includegraphics[width=0.55\textwidth]{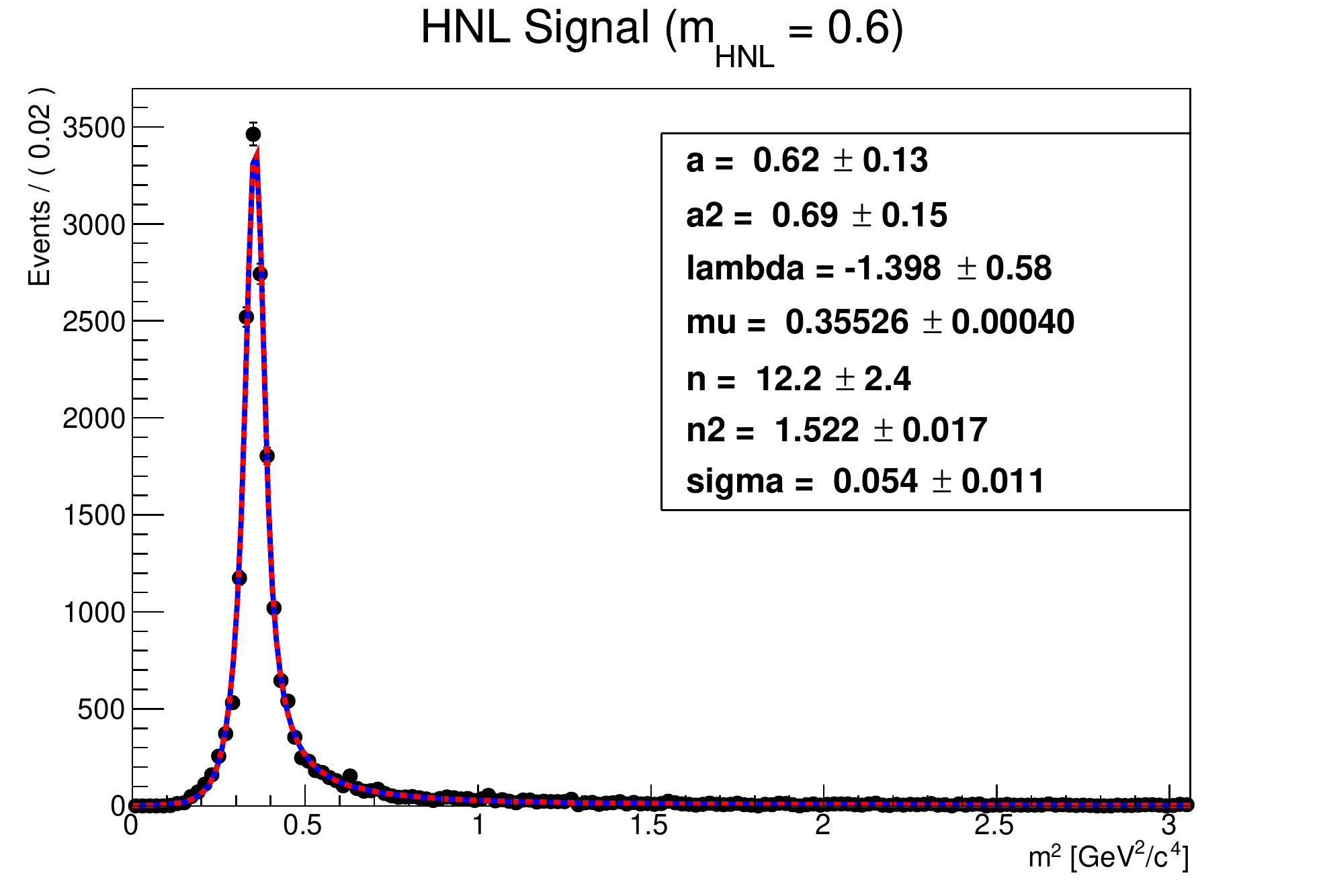}}\\
    \subfloat[]{\includegraphics[width=0.55\textwidth]{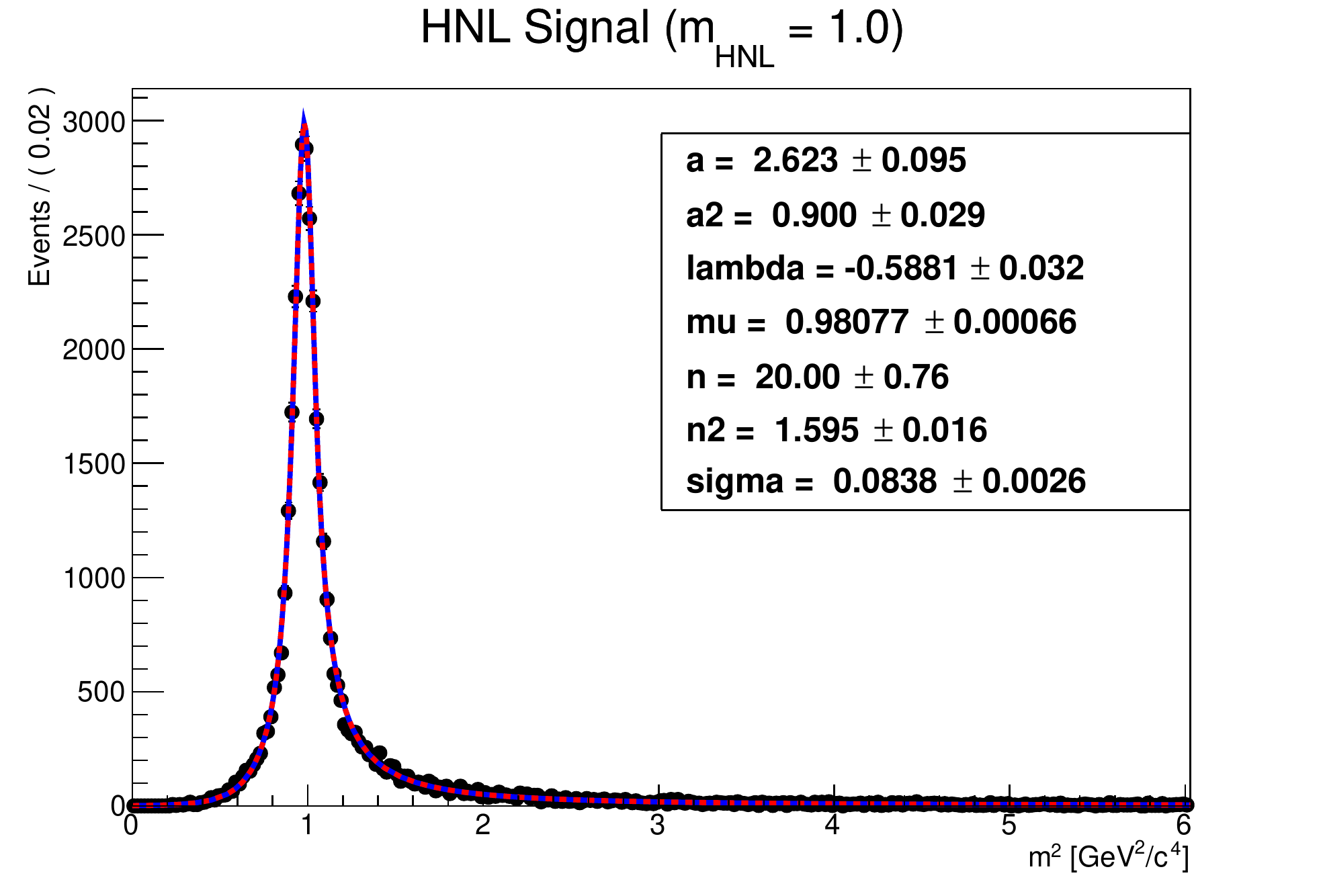}}
    \subfloat[]{\includegraphics[width=0.55\textwidth]{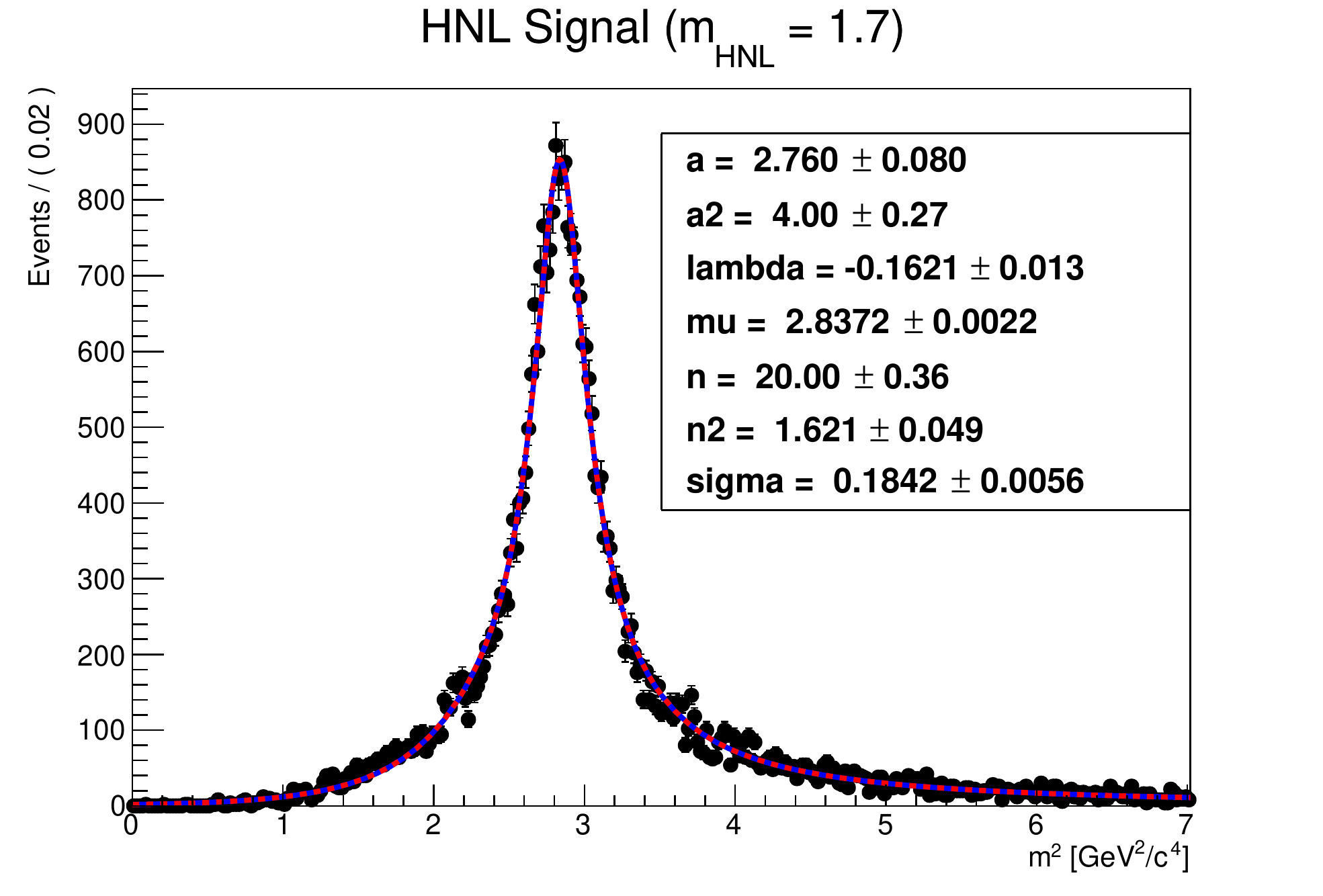}}
    \caption{\small{The invariant mass of the signal model for accepted events in $D_s\rightarrow \mu N_2$, for (a) $M_N=0.3$ GeV/$c^2$, (b) $M_N=0.6$ GeV/$c^2$, (c) $M_N=1.0$ GeV/$c^2$, (c) $M_N=1.8$ GeV/$c^2$. For better visual (a) is in log scale.}}
    \label{fig:invmass}
\end{figure}

The HNL signal candidate is defined as a selected track pair used to evaluate the invariant mass. The signal distribution is modeled with two sided Hypatia p.d.f.\cite{hypa:2014}\footnote{Hyperbolic core of a crystal-ball-like G function and two tails, used in modeling invariant mass distribution with generic tails. The parameters of the function like $\mu$ and $\sigma$ represent the mean and sigma of the gaussian component, while the other parameters are needed to deal with the tails} using RooFit. 
\subsection{Background}
The most generic background for this signal is the neutrino interactions from the beam for six years of exposure. Considering a single beam spill, with the period of $1.2 s$, the total number of neutrino interactions, for six years of exposure, within the full body of the detector is around $1.3\times10^{10}$, which is not compatible with the available computational power. Consequently, to have an estimate on the background, few assumptions have been made:

\begin{enumerate}
\item Neutrino CC  interactions, only inside the SAND inner tracker (STT): $\sim 10^8$
\item Keeping the high statistics only at the generation level, selecting the most dangerous background ($\nu_{\mu}$ CC interaction with single $\pi$ at final state): 30$\%$ of the total events
\end{enumerate}

\begin{figure}[!htbp]
    \centering
    \includegraphics[width=0.55\textwidth]{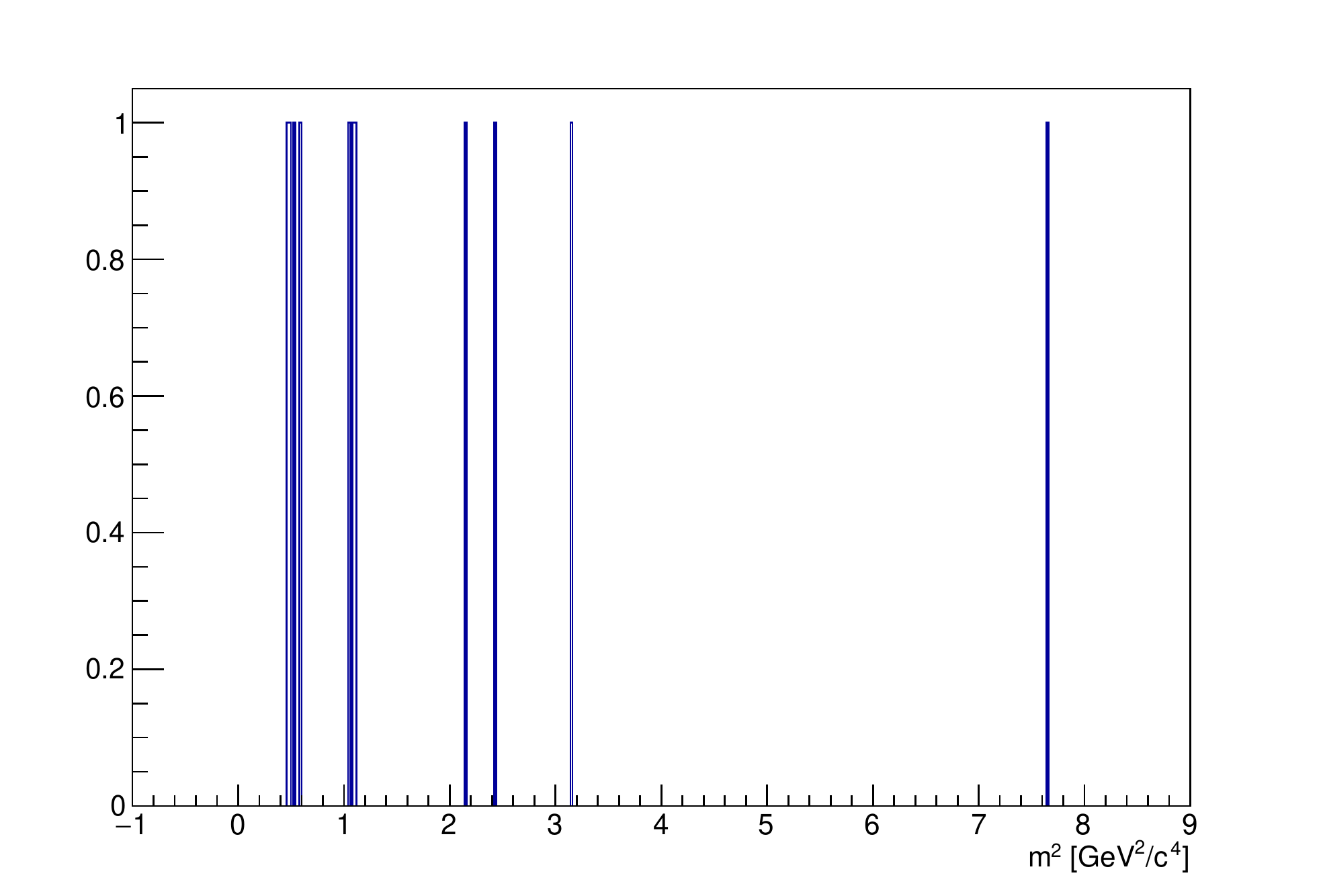}
    \caption{\small{The distribution of invariant mass from the cherry picked background of $\nu$ CC interactions with one $\pi$ at final state, emulating the signal}}
    \label{fig:bkginv}
\end{figure}

The first assumption seems logical, since the signal is inside the inner tracker, but the statistics is still too high to manage. Potential additional background from interactions outside of the STT can be rejected by the selection, possibly with minor reduction of the fiducial volume. By adding the second assumption, taking into account the most dangerous background, $\pi\mu$X\footnote{The requirement for 4 tracks (phys+ghosts) and a vertex within 1mm makes other background topologies sub-dominant, if not negligible, in terms of relative contamination; combined with the relative abundance (~30$\%$) of the selected background sample makes other contributions negligible}, carried out by using the cherry picking technique of the neutrino event generator, GENIE\cite{Andreopoulos:2009zz}, reduces the statistics to 30$\%$, which is manageable.
 This pre-selected background is passed through the complete chain as the signal simulation and reconstruction.
The resulting background invariant mass distribution (Fig. \ref{fig:bkginv}) is limited to 11 final background candidates in six years of exposure; although it exhibits a potential preference for lower values, it is not possible to univocally select a model to represent it, especially because in the signal region ($[0,3.5]$ GeV$^2$/$c^4$) it appears to be uniform.
Hence, both a uniform and an exponential p.d.f. \footnote{probability distribution function}model is considered.

\subsection{Final Sensitivity}

The final sensitivity is estimated by combining the method and parameters used for Pheno-sensitivity with the statistical analysis of the signal and background distributions after the selection.
Around 100 toy MC is generated, based on the signal and background models, in order to calculate the confidence level (CL). The statistical technique in calculating the CL has been carried out by RooFit, following a frequentist approach, based on the likelihood ratio.
 
\begin{figure}[!htbp]
    \centering
    \subfloat[]{\includegraphics[width=0.5\textwidth]{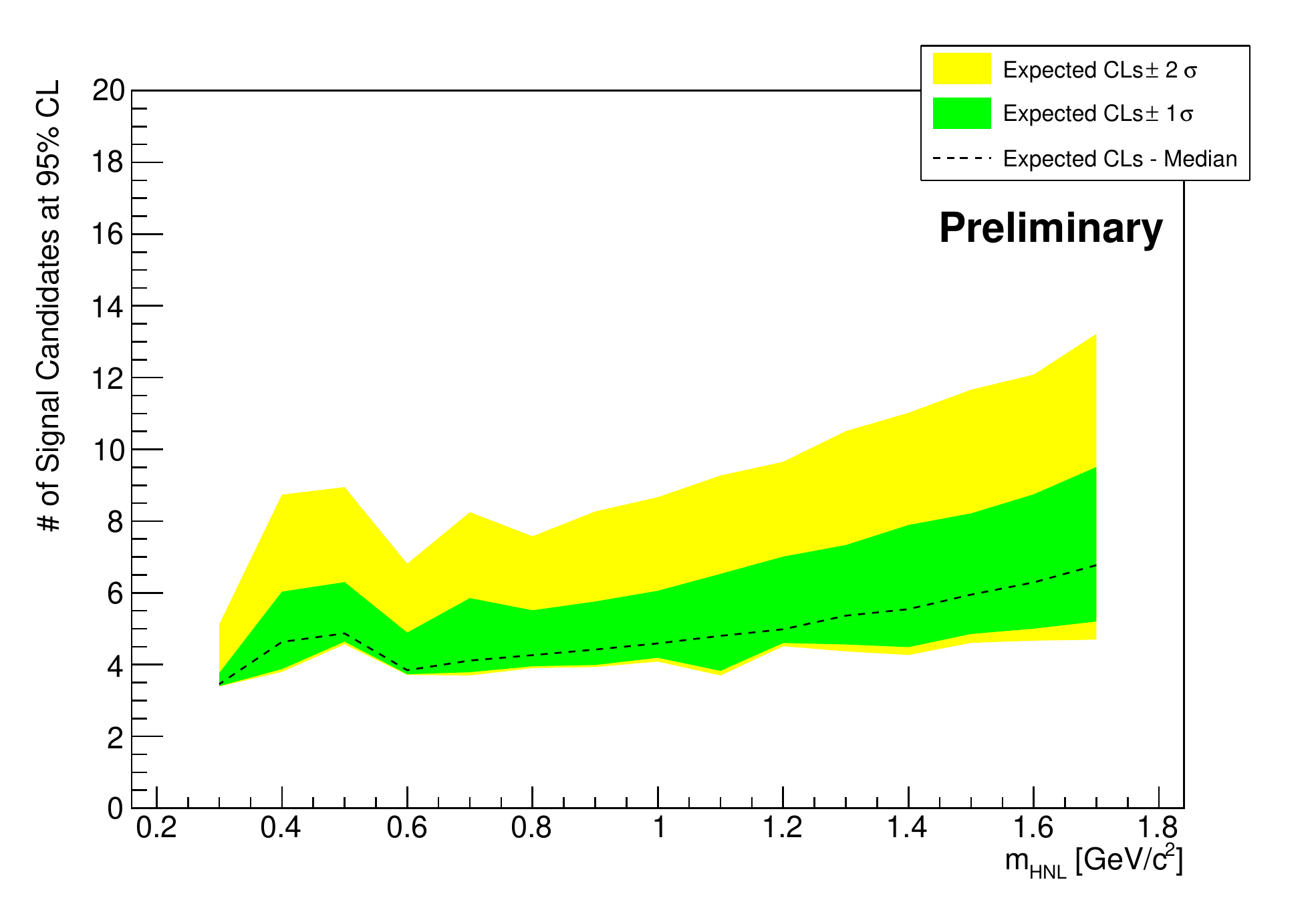}}
    \subfloat[]{\includegraphics[width=0.5\textwidth]{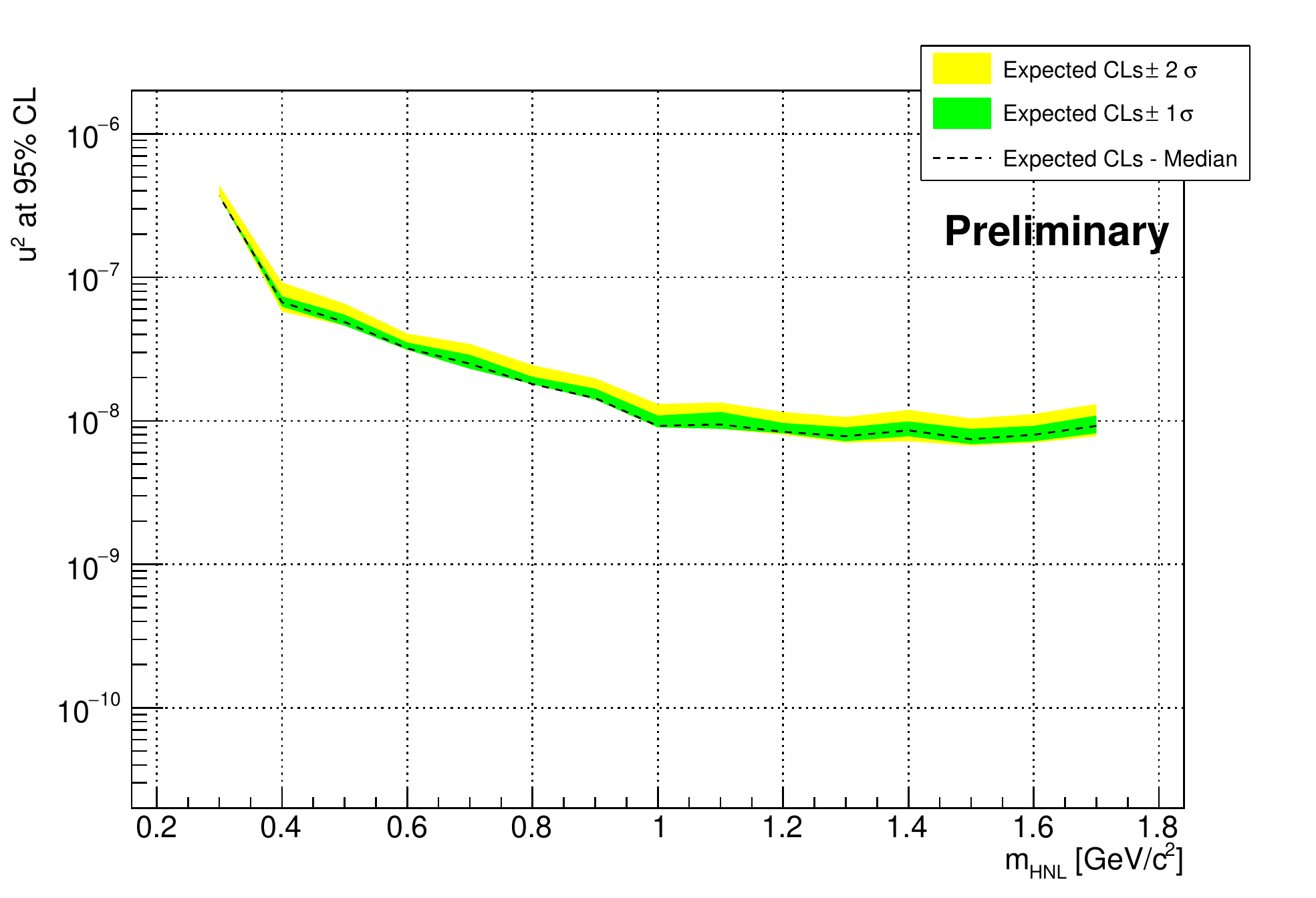}}
    \caption{\small (a) Number of signal candidates corresponding to 95\% CL for each HNL mass, (b) Final sensitivity to HNL searches in SAND at 95\% CL for benchmark II (Majorana HNL assumption)}
    \label{fig:sensN}
\end{figure}
Comparing the final sensitivity with the Pheno-sensitivity, it can be observed that the introduction of the experimental setup over the simple assumption of full efficiency and zero background weakens the performance, but only by a factor $\approx 3$. This is the combination of a slight overestimate of the detector volume, using a cube in the Pheno-sensitivity, and the global reconstruction and selection efficiency.

\section{Conclusion}

This work elaborates an investigation of SAND sensitivity to HNL, in a minimally extended Standard Model scenario, $\nu$MSM. Starting with the lagrangian, the workflow of this study goes through several connected steps that, once concatenated, allow to estimate the final sensitivity.
\begin{equation*}
\text{\textbf{Model}} \Rightarrow \text{\textbf{MC-Gen}} \Rightarrow \text{\textbf{Det-Res}} \Rightarrow \text{\textbf{Digit}} \Rightarrow \text{\textbf{Reco}} \Rightarrow \text{\textbf{Sel}}
\end{equation*}

Including $\nu MSM$ parameters and coupling benchmarks I,II,III, the Lagrangian is translated into a Monte Carlo compatible code, through FeynRules; it is, then, used in a modified version of the Mad-Graph5, Mad-Dump, MC event generator.
There are two MC generators used in this simulation, Mad-Dump as the primary and Pythia8 as the auxiliary. 

For reconstruction, a customized Kalman Filter (KF) has been implemented, designed for the purpose of this study, for DUNE-ND SAND with the prospect of extended use in a more generalized context. The performance of KF is already quite good, given the high estimated efficiency.
A selection, based on few cuts on the angle between the oppositely charged tracks, removes all ghosts that would degrade the invariant mass resolution, leaving a harmless yield contamination that is evaluated and added as a normalization factor.
The invariant mass for signal candidates is modeled with an appropriate p.d.f., two-sided Hypatia distribution function. The estimated background shows a few candidates mimicking the signal over six years of exposure; the sample is statistically limited, therefore it is modeled either with a uniform or exponential p.d.f., demonstrating very similar results.

Comparing the phenomenological and final sensitivities a small degradation, around a factor $\approx 3$, is observed; considering that the Pheno-Sensitivity is estimated on a detector volume slightly larger than the actual one, it is clear that reconstruction and selection are performing already very well. A margin for future improvement, both in efficiency and background rejection, would probably allow to retain the current performance in more realistic conditions rather than improving on the current result.

The procedure and tools developed for this study can easily accommodate the inclusion of the other meson families, which would result in an improved sensitivity; nevertheless, being the $D_s$ a major contributor, the current result is not expected to be very far from what it would be after including all the other mesons in the mass range $[0.5,1.8]$ GeV/$c^2$.

The estimated final sensitivity improves on existing limits, by orders of magnitude for the higher masses, and is competitive with expectations from other foreseen projects. Extending the study to other detectors in the DUNE-ND complex, with larger volume, would improve the current result, provided that comparable signal detection efficiency and background rejection could be achieved.


\end{document}